\documentclass[twocolumn]{aastex631}

\usepackage{CJKutf8, color}

\usepackage{url,xcolor,CJKutf8,acronym,ulem,braket,hyperref,amsmath}

\usepackage{graphicx}
\usepackage{multirow}

\begin{document}

\shorttitle{Evndence of 30-keV Hump on \textit{rp}-process}
\shortauthors{Kubota, Dohi et al.}

\title{Hump around 30 keV in \textit{Suzaku} spectra of Aquila X-1\\ -- Possible evidence of strong rapid-proton capture process --}

\correspondingauthor{Megu Kubota}
\email{megu.kubota@phchd.com}

\correspondingauthor{Akira Dohi}
\email{akira.dohi@riken.jp}

\author[0000-0001-7298-3243]{M.~Kubota }
\affiliation{High Energy Astrophysics Laboratory, RIKEN, Wako, Saitama 351-0198, Japan}
\affiliation{Department of Physics, Tokyo University of Science, 1-3 Kagurazaka, Shinjyuku-ku, Tokyo 162-0825, Japan}
\altaffiliation{Current affiliation: PHC corporation, 15th Floor, Dai-ichi Life Hibiya First, 1-13-2 Yurakucho, Chiyoda-ku, Tokyo 100-0006, Japan}

\author[0000-0001-8726-5762]{A.~Dohi }
\affiliation{Interdisciplinary Theoretical and Mathematical Sciences (iTHEMS), RIKEN, Wako, Saitama 351-0198, Japan}

\author[0000-0002-8801-6263]{T.~Tamagawa }
\affiliation{High Energy Astrophysics Laboratory, RIKEN, Wako, Saitama 351-0198, Japan}
\affiliation{Department of Physics, Tokyo University of Science, 1-3 Kagurazaka, Shinjyuku-ku, Tokyo 162-0825, Japan}

\author[0000-0002-1040-8769]{K.~Makishima }
\affiliation{Department of Physics, The University of Tokyo, 7-3-1 Hongo, Bunkyo-ku,Tokyo 113-0033, Japan}
\affiliation{Kavli Institute for the Physics and Mathematics of the Universe (WPI), The University of Tokyo, 5-1-5 Kashiwa-no-ha, Kashiwa, Chiba 277-8683, Japan}

\author[0000-0002-7273-2740]{S.~Fujimoto }
\affiliation{National Institute of Technology, Kumamoto College, Kumamoto 861-1102, Japan}

\begin{abstract}

In a neutron star low-mass X-ray binary, heavy elements (with mass number $\sim$100) 
are believed to be synthesized through the rapid-proton (rp) capture process, as indirectly indicated by the long tail of X-ray burst light curves. Although direct evidence for the \textit{rp}-process was lacking so far,
we reported in our previous study \citep{paper1}
that an unidentified hump is present at $\sim 30$ keV of
the \textit{Suzaku} spectra of Aquila X-1 in its hard state.
The feature was tentatively identified with a recombination quasi-continuum 
from highly ionized ions of cadmium or indium,
which are possibly produced by the \textit{rp} process during X-ray bursts
and lifted up into the neutron-star atmosphere.
To reinforce this scenario, we quantitatively evaluated the physical conditions of the assumed emitting region in the atmosphere,
and confirmed that the scenario is possible
when the putative \textit{rp}-process ashes in the atmosphere
has $\sim 10^7$ times higher abundances than their solar values.
The results are compared with theoretical model calculations for the nucleosynthesis in X-ray bursts. 
We suggest that the observed spectral hump is
a radiative recombination feature from
the stable isotope $^{106}{\rm Cd}$, 
produced in X-ray bursts via the \textit{rp}-process. Further issues about the detectability are briefly discussed.
\end{abstract}



\section{Introduction}
\label{Introduction}

When a neutron star (NS) with a low magnetic field forms
a mass exchanging binary with a low mass companion star,
a system called a low mass X-ray binary (LMXB) is formed.
These systems often emit Type-I X-ray Bursts (XRBs),
when hydrogen and helium in the accreting matter undergo
unstable thermonuclear burning on the NS surface.
Unless the donor is a helium star, the fresh accreting matter in LMXBs consists mainly of hydrogen.
Therefore, these bursts provide a promising site 
for the rapid proton-capture (rp) process;
a sequence of proton captures, interleaved with $\beta^+$ decays,
synthesize proton-rich heavy elements
that are far from stability and beyond the iron group
(for reviews, see e.g., \cite{2006NuPhA.777..601S,2013PrPNP..69..225P}). According to a pioneering network calculation by \cite{Schatz01},
the \textit{rp} process has a natural endpoint at a closed SnSnTe cycle.

The heating by the \textit{rp}-process products is expected to provide sustained energy release during a burst decay phase, 
and imprint characteristic signatures in the cooling tail
\citep{Woosley_2004,Fisker:2004sc}. This effect is mainly caused by waiting-point nuclei, which slow down the reaction chain 
due to relatively slow $\beta^+$-decays  or inhibited proton captures. Indeed, systematic analysis of a large sample of X-ray bursts has revealed 
that many hydrogen-rich bursts exhibit additional tail components that are consistent with the \textit{rp}-process heating \citep{2017A&A...606A.130I}. The general presence of this component is also suggested
by the characteristic timescale of Type I XRBs,
which is typically $\sim40~{\rm sec}$ in hydrogen-rich burst sources,
whereas $\lesssim20~{\rm s}$ in
helium-rich ones~(e.g., \cite{2025PASJ...77L..24T}).

Despite the strong theoretical motivation and the indirect evidence from the burst light curves, 
direct observational signatures of \textit{rp}-process ashes remain rare at present, although some reports have suggested the presence of photoionization edges or absorption features associated with heavy elements during bursts (e.g., in HETE J1900.1-2455; \citet{10.1093/mnrasl/slw167}). Therefore, firmer detections of spectral features from the \textit{rp}-process products remain one of the major open challenges in NS astrophysics. 

In our previous study (\cite{paper1}; hereafter Paper I), 
we analyzed broadband \textit{Suzaku} spectra of Aquila X-1 (Aql X-1 for short), one of the representative LMXBs, acquired during a decay phase of its 2007 outburst. At $\sim 30$ keV of the spectra, we noticed a local excess feature,
which remained significant against various modelings of the LMXB continuum spectrum. 
The excess was tentatively interpreted as a recombination-edge (\texttt{redge}) feature, 
from heavy elements (Cd, In) that might have been synthesized via the {rp}-process. 
In the present paper, we improve the \textit{Suzaku} spectral analysis, 
and examine carefully whether this interpretation is feasible,
in view of various physical processes in the NS atmosphere, assuming that the heavy elements inferred from the observations
are explained by the theory of XRBs. By doing so, we attempt to achieve one of the first observational confirmations 
of the \textit{rp} process taking place in LMXBs.

This paper is organized as follows. Section~\ref{section:paperI} revisits the 30 keV feature discovered in Paper I, and presents its refined analysis.
In section~\ref{section:constraints}, the observed information is converted into requirements on the amount and physical conditions 
of the \textit{rp}-process products (Cd, In),
which are then compared in section~\ref{sec:xrbcalc} with predictions from the theory of \textit{rp}-process nucleosynthesis. 
After identifying remaining issues in section~\ref{sec:further_issues}, 
we conclude in section~\ref{section:conc}.

\section{The Spectral feature around 30~keV}
\label{section:paperI}

Here we review the key results derived in Paper I. Aquila X-1 was observed seven times in total with \textit{Suzaku},
during the decay phase of an outburst in 2007.
Table \ref{tab:suzakuobs} summarizes
the first four observations (September 28--October 15).
%
\begin{table}[h]
    \caption{\textit{Suzaku} observations of Aql X-1 in 2007 (see \cite{Sakurai12} for details).}
    \label{tab:suzakuobs}
    \centering
    \begin{tabular}{cccccccc}
        \hline\hline
        No. && Date  && Obs.ID    && State   &\\ \hline
        D1  && 9/28  && 402053010 && Soft    & \\ 
        D2  && 10/03 && 402053020 && Hard    &\\ 
        D3  && 10/09 && 402053030 && Hard    &\\ 
        D4  && 10/15 && 402053040 && Hard    &\\ \hline
    \end{tabular}    
\end{table}
The source was in a typical soft state
in the first observation (D1),
whereas it was in a hard state in the subsequent 
three observations, D2, D3, and D4.
The broadband (1--100 keV) spectra from the three hard-state observations were very hard,
and all exhibited a pronounced excess feature at $\sim 30$ keV. 
When modeled with a Gaussian   
added on top of an empirical continuum model (see below),
this feature had an Equivalent Width (EW) of $7.3_{-3.2}^{+3.2}$ keV, 
$8.8_{-3.5}^{+6.4}$ keV, and $3.7_{-2.5}^{+2.6}$ keV in D2, D3, and D4, respectively.

\begin{figure}[t]
\vspace{1cm}
\centering    \includegraphics[width=\linewidth, trim=150 250 150 210, clip]{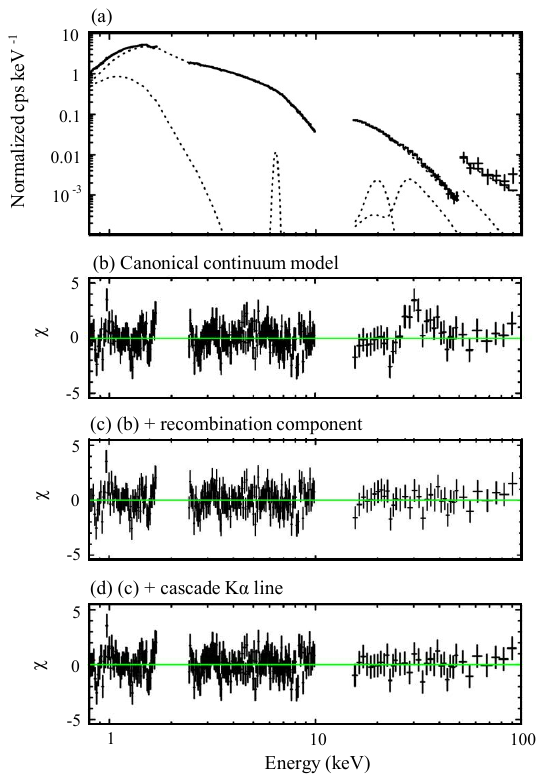}
\centering
\caption{
(a) The combined D2--D4 spectrum,
shown together with the best-fit model that 
includes a \texttt{redge} component 
and an additional Gaussian 
for the cascade K$\alpha$ line at $\sim 20$~keV.
(b) Residuals from the canonical continuum model (for reference).
(c) Same as (b) but with the \texttt{redge} component added to the model.
(d) Fit residuals from the best-fit model shown in panel (a).
}
\label{fig:cascade}
\end{figure}

In Paper I, we combined the spectra from D2, D3, and D4 
into a summed spectrum reproduced in figure~\ref{fig:cascade}.
It was fitted by various continuum models,
including the canonical one which has a form of
\texttt{tbabs * \{bbody + CompPS(bbody) + Gaussian(FeK)\}}.
Here, \texttt{tbabs} denotes the foreground interstellar absorption~\citep{1996ASPC..101...17A},
with the Solar abundance~\citep{2000ApJ...542..914W} and an quivalent hydrogen column density of $N_{\rm H}=0.36\times10^{22}~{\rm cm}^{-2}$ (see also Section 3.1 in Paper I). \texttt{CompPS(bbody)}~\citep{1996ApJ...470..249P} is a thermal Comptonization continuum
from the hot and optically-thin accretion flow,
for which the seed-photon source is assumed to be 
a blackbody emission from the NS surface, represented by \texttt{bbody}. 
A certain fraction of the blackbody emission is assumed to thus undergo Comptonization in the hot corona, whereas the rest is directly visible 
at the lowest end of the spectrum. 
A low-ionization Fe-K emission line at 6.4 keV, presumably from outer regions of the accretion disk, is modeled by \texttt{Gaussian(FeK)}.

In Paper I, we confirmed that the 30 keV structure is significant,
both statistically and systematically,
when evaluated against the canonical spectral model described above.
This condition is shown in figure~\ref{fig:cascade}b,
where the feature appears as residuals at $\sim 30$ keV. Furthermore, as in panel (c) of the same figure, the feature in the residuals was well explained away by a radiative recombination (\texttt{redge}) model, which describes a process
that free-bound emission from highly ionized ions forms a quasi continuum above their K-edge. The best-fit K-edge energy was $E_{\rm K}=27 \pm 1$ keV, and the electron temperature was $kT_{\rm e}=11^{+10}_{-5}$ keV 
as determined by the quasi-continuum width.
After correcting for the gravitational redshift 
at the surface of a typical NS~\footnote{For the general-relativistic correction, see e.g., Section 3.2 in \cite{JOHNSTON2020}.}, the inferred edge energy 
implies heavy elements with atomic numbers 
$Z \approx 48$-49, namely, Cd or In.

Physically, a \texttt{redge} component should be accompanied by cascade emission lines, 
which were not considered in Paper I. The strongest among them is the K${\alpha}$ transition, 
expected at $\simeq 0.75\,E_{\rm K}\approx20$ keV in our case.
To quantify its expected intensity, we assumed a fully ionized Cd/In plasma with $kT_{\rm e}=11$ keV, 
and calculated the free-bound transition coefficients 
into the 2p levels and into the ground state.
The results were obtained as
$4.33\times10^{-15}$ s$^{-1}$ and $1.76\times10^{-14}$ s$^{-1}$, 
respectively \citep{2008CaJPh..86..675G}.
This predicts a K$\alpha$ cascade line with an intensity 
roughly one-fourth of the recombination continuum.

To examine the data for the above expectation, 
we repeated the spectral fitting by adding 
a narrow Gaussian line at $\sim$20 keV to the 
successful model in Paper I (the canonical model + \texttt{redge} component).
As shown in Figure \ref{fig:cascade}d, 
the fit was slightly improved from 
$\chi^2/{\rm dof}=220.56/216$ to $216.89/214$, 
while the continuum  and \texttt{redge} parameters were almost unchanged.
In particular, the significance of the 30 keV hump remained unaffected.
The Gaussian energy was obtained as $20^{+1}_{-2}$ keV 
with a marginally positive flux of 
$(8.3^{+7.3}_{-7.0})\times10^{-5}$ photons cm$^{-2}$ s$^{-1}$.
Compared with the \texttt{redge} flux of 
$(6.0^{+2.7}_{-2.0})\times10^{-4}$ photons cm$^{-2}$ s$^{-1}$, 
this means a flux ratio of $0.14^{+0.24}_{-0.02}$, 
which is consistent with the expected value of $\sim 0.25$.
Thus, the data accommodate a cascade K$\alpha$ line whose intensity is consistent with the theoretical prediction.

\section{Astrophysical Constraints from Observations}
\label{section:constraints}

We have successfully explained the overall spectrum 
with a reasonable model, 
which consists of a canonical continuum model,
and \texttt{redge} plus K$\alpha$ components attributed to 
$Z \simeq 48$ (Cd or In) elements in the atmosphere. 
However, the observed \texttt{redge} feature has such 
a large EW reaching 5.6 keV,
that the relevant elements must be enriched by many orders of magnitude above their solar abundances ($6\times10^{-11}$ for Cd). 
Thus, the required amount of Cd/In enrichment must be quantified.
Furthermore, such an atmosphere with exceptionally high metallicity, 
if realized at all,
may behave differently from the ordinary hydrogen-rich ones.
Hence, below we identify overall astrophysical requirements 
for this scenario to become valid, 
and examine whether we can find an atmospheric condition
that consistently satisfies all these observational requirements.

\subsection{Basic assumptions}
\label{subsection:assumptions}

To begin with, we assume the following conditions for the NS atmosphere.
\begin{enumerate}
\setlength{\itemsep}{0cm}
\item The atmosphere consists solely of electrons, protons, and Cd/In ions,
which have densities $n_{\rm e}$, $n_{\rm H}$, and $n_{\rm x}$, respectively, 
 at the height above the NS surface 
 where the \texttt{redge} feature is produced.
\item The average ion mass is given by
\begin{equation}
m_{\rm i} = m_{\rm p} \left(\frac{A_{\rm x} n_{\rm x}+n_{\rm H}}{n_{\rm x}+n_{\rm H}} \right),
\label{eq:mi}
\end{equation}
where $m_p$ is the proton mass, and $A_{\rm x}\sim 106$ is 
the most abundant mass number of Cd/In isotopes (see section~\ref{sec:xrbcalc}).

\item The ions are fully ionized in 
the relevant region.
Thus, the charge neutrality requires
\begin{equation}
n_{\rm e} = 48 n_{\rm x} + n_{\rm H}  ~. 
\label{eq:ne}
\end{equation}

\item
The atmosphere is isothermal,
with a temperature of $kT_{\rm e} \simeq 11$~keV
as indicated by the \texttt{redge} fit (section~\ref{section:paperI}),
and has an exponential density distribution 
with a scale height of
\begin{equation}
H = \frac{k T_{\rm e}}{m_{\rm i}g }
\simeq  \frac{m_{\rm p}}{m_{\rm i} }  \times {66}~~{\rm cm}
= \frac{n_{\rm x}+n_{\rm H}}{A_{\rm x} n_{\rm x}+n_{\rm H}} \times {66}~~{\rm cm}.
\label{eq:scaleheight}
\end{equation}
Here, $g=1.6\times10^{14}~{\rm cm~s^{-2}}$ 
is the gravitational acceleration on the NS surface, 
assuming the NS mass and radius of $1.4\,M_\odot$ 
and $R_{\rm NS}=12~\mathrm{km}$, respectively. 
The last expression utilizes equation~(\ref{eq:mi}). 
\end{enumerate}

\begin{figure}[t]
\centering
\includegraphics[width=\linewidth, trim=135 70 150 0, clip]{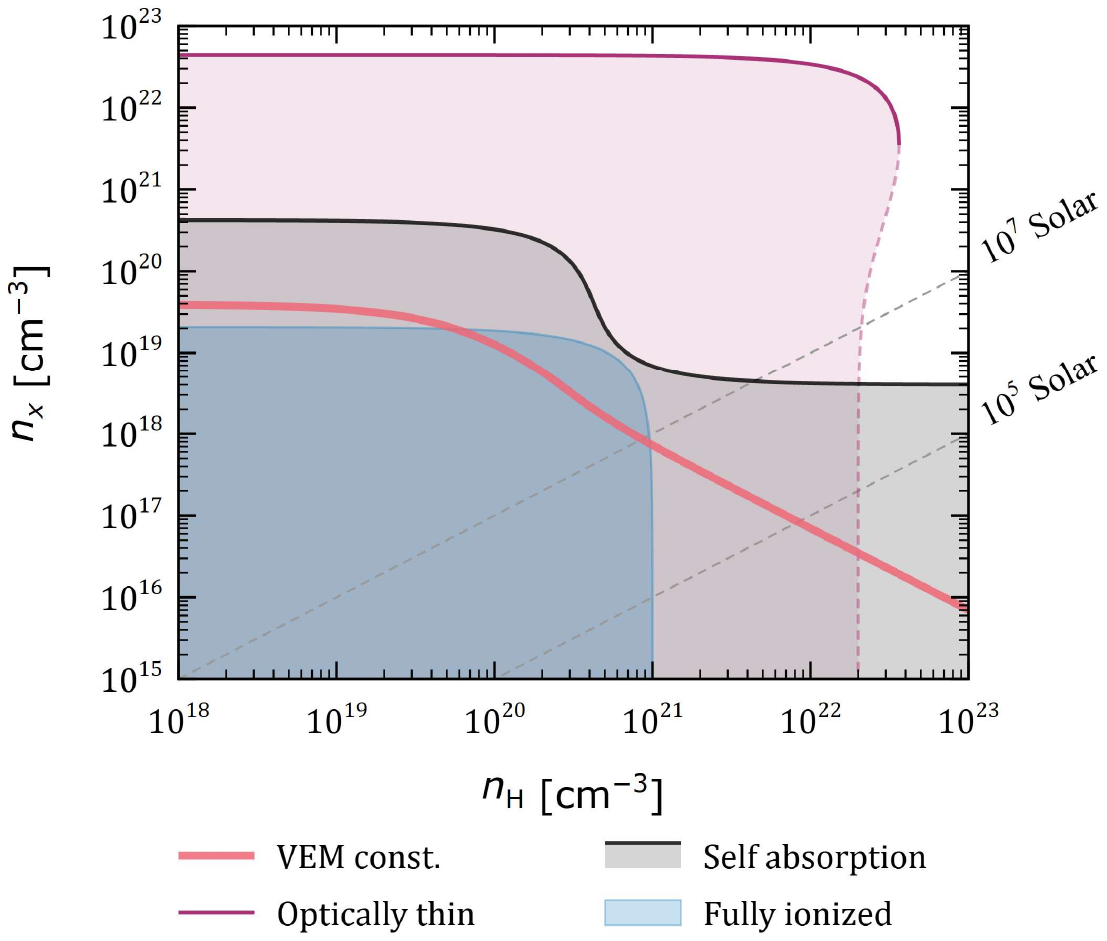}
\caption{
A diagram of hydrogen number density, $n_H$, versus that of Cd, $n_x$, 
in the NS atmosphere.
A pair of dashed gray lines shows $10^7$ and $10^5$ times 
the Solar abundance of Cd. 
The red curve indicates the VEM of equation~(\ref{eq:VEM}) 
which is needed to explain the observed flux of the recombination photons. 
The blue shaded region is allowed by the condition for high ionization [equation~(\ref{eq:17})]. The magenta curves and shades indicate the constraints 
by the Thomson optical depth [equation (\ref{eq:19})]. The black curve and gray shades indicate the constraints 
by the self-absorption [equation~(\ref{eq:21 condition self abs})], respectively. 
}
\label{fig:conditions}
\end{figure}

\subsection{Required amount of heavy elements}
\subsubsection{Volume emission measure}
\label{subsubsec:vol_emission_measure}
%
On the $(n_{\rm H}, n_{\rm x})$ plane in figure~\ref{fig:conditions},
let us calculate the expected intensity of the \texttt{redge} feature.
Generally, the number of recombination photons emitted from
a recombining plasma with a volume $V$ is given as 
\begin{equation}
\Phi_{\rm rec} = n_{\rm x} n_{\rm e} V 
\int_{E_{\rm K}}^{+\infty} S(E)dE ~
\label{eq:num_photon}
\end{equation}
per unit time in all energies $E$
above the ionization potential $E_{\rm K}$.
The quantity $n_{\rm x} n_{\rm e} V$ is 
called the volume emission measure (VEM), 
and the function $S(E)$, having a dimension of 
photons$\cdot$ volume time$^{-1}$ energy$^{-1}$, is given by \mbox{\citep{SRON_rrc}} 
\begin{equation}
\begin{split}
S(E) dE&=\sqrt{ \frac{2}{\pi}}
\frac{g_{\rm n} \sigma_{\rm bf} E^2 }{(m_{\rm e}kT)^{3/2}c^2}
\exp{\left(\frac{E_{\rm K}-E}{kT} \right)} dE\\
&=\sqrt{\frac{2}{\pi}} g_{\rm n} 
\frac{\sigma_{\rm bf} c x^2}{(a\,b)^{3/2}}
\exp{\left(\frac{1-x}{a} \right)} dx
\end{split}
\label{eq:emissivity}
\end{equation}
where $c$ is the light velocity,
$m_{\rm e}$ is the electron mass,
and $g_{\rm n} = 2$ is the statistical weight 
of the recombining level ({\it 1s} level in the present case),
while $x \equiv E/E_{\rm K}$, 
$a \equiv k T_{\rm e}/E_{\rm K}=0.407$
(assuming $k T_{\rm e}=11$ keV),
and $b \equiv m_{\rm e} c^2/E_{\rm K} = 18.9$
are dimensionless parameters.
The recombination cross section $\sigma_{\rm bf}$ 
can be written as \mbox{\citep{SRON_rrc}}
\begin{equation}
\sigma_{\rm bf} = 
\left(\frac{A g_{\rm bf}}{n} \right)
\left(\frac{E_{\rm K}^2}{E^3}\right)
= \left(\frac{A g_{\rm bf}}{n} \right)\frac{1}{ E_{\rm K} x^3}
~{\rm cm^2}, 
\label{eq:sigma_bf}
\end{equation}
where 
$A = 1.1\times 10^{-19}$ cm$^{2}$ keV is a constant,
$n$ is the principal quantum number 
(=1 in the present case),
$g_{\rm bf} \simeq 0.2$ is the Gaunt factor \citep{1961ApJS....6..167K}, 
and $E_{\rm K}$ is in units of keV.

Combining equation~(\ref{eq:emissivity}) and equation~(\ref{eq:sigma_bf}), 
we obtain, in units of ${\rm cm^{3}~s^{-1}~keV^{-1}}$,
\begin{equation}
\begin{split}
S(E) dE &= \sqrt{\frac{2}{\pi}} 
\frac{g_{\rm n} g_{\rm bf}}{n(a\,b)^{3/2}}
\left(\frac{A\,c}{E_{\rm K}} \right) 
\frac{1}{x}\exp{\left(\frac{1-x}{a} \right)} dx\\
&=1.8 \times 10^{-12} 
\frac{1}{x}\exp{\left(\frac{1-x}{a} \right)} dx
\end{split}
\label{eq:emissivity2}
\end{equation}
where the actual values of the various parameters were employed.
Substituting this into equation~(\ref{eq:num_photon}),
we obtain
\begin{equation}
\Phi_{\rm rec} \sim n_{\rm x} n_{\rm e} V 
\times 10^{-12} ~{\rm ph~cm^{3}~sec^{-1}}.
\end{equation}
A more rigorous calculation by \citet{Badnell} gives 
\begin{equation}
\Phi_{\rm rec} =n_{\rm x} n_{\rm e} V 
\times 5 \times 10^{-12} ~{\rm ph~cm^{3}~sec^{-1}}.
\label{eq:num photon3}
\end{equation}
which is adopted hereafter. 

In our observations, 
the recombination flux $F(E)$ was measured at the Earth as 
\begin{equation}
\Phi/4\pi D^2 = F(E) = 5.0 \times 10^{-4} ~{\rm ph ~cm^{-2} s^{-1}}, 
\end{equation}
where $D=5.2$~kpc is the distance to the source 
\citep{AqlX-1_Dist} \footnote{This is very close to that published in the data release 3 of Gaia observations, $D=5.55\pm3.57$~kpc~\citep{2021A&A...649A...1G,2022MNRAS.510.1577G} (but see also the value of data release 2, $D=2.97^{+2.64}_{-1.32}$~kpc~\citep{2018A&A...616A...1G,galloway20}). These distance uncertainties do not change our conclusion.
}. 
Therefore, the data require
\begin{equation}
\Phi = 4 \pi F(E) D^2 = 1.6 \times 10^{42} ~{\rm ph~s^{-1}} ~. 
\label{eq:num photon obs}
\end{equation}
By combining equation~(\ref{eq:num photon3}) and 
equation~(\ref{eq:num photon obs}), 
the required VEM can be obtained as 
\begin{equation}
n_{\rm x} n_{\rm e}V \sim 3 \times 10^{53} ~{\rm cm^{-3}} ~.
\end{equation}
Assuming that half the atmosphere is occulted by the NS, 
the emitting volume in the atmosphere can be expressed as 
$V = 2\pi R_{\rm NS}^2 H = 9 \times 10^{12} H~{\rm cm^3}$. 
Thus, we finally obtain
\begin{equation}
n_{\rm x} n_{\rm e} H \sim 5 \times 10^{40} ~{\rm cm^{-5}} ~. \label{eq:req condition1}
\end{equation}

Using equation~(\ref{eq:ne}) and equation~(\ref{eq:scaleheight}), 
the constraint of equation~(\ref{eq:req condition1}) can be rewritten as 
\begin{equation}
\frac{n_{\rm x} (n_{\rm H} + n_{\rm x})(n_{\rm H} + 48 n_{\rm x})}{n_{\rm H} + A_{\rm x} n_{\rm x}} = 7 \times 10^{38} ~{\rm cm^{-6}}.
\label{eq:VEM}
\end{equation}
which is indicated in figure~\ref{fig:conditions} by the red line.
Despite the complex functional form, 
its asymptotic behavior can be easily understood;
it reduces to
\begin{equation}
n_{\rm x} \sim (7 \times 10^{38} A_{\rm x}/48)^{1/2} \sim 4\times 10^{19}~{\rm cm}^{-3}
\end{equation}
for $n_{\rm x} \gg n_{\rm H}$ 
where electrons are mostly supplied by the heavy ions,
whereas for $n_{\rm H} \gg A_{\rm x}n_{\rm x}$  we find
\begin{equation}
n_{\rm H}\, n_{\rm x} \sim 7\times 10^{38} ~{\rm cm}^{-3}
\end{equation}
because electrons mainly come from hydrogen.

\subsection{Ionization of heavy elements}
For the heavy-element ions to produce the \texttt{redge} emission, they must be nearly fully ionized, requiring that their ionization temperature is much higher than that of the ambient electrons (11 keV). Such a condition can be realized when the ions are photoionized,
irradiated by photons that have a color temperature exceeding 
$E_{\rm K} \approx 27$ keV.
This necessary condition was basically realized 
in the D2 to D4 observations,
because the source was then in the hard state,
and the thermal Compton X-rays from the inflowing matter
had a sufficiently high temperature of $\sim 55$ keV (Paper I).

Even when the illuminating X-rays have a 
high temperature as confirmed above,
the heavy-element ions would not become fully ionized unless
the ionization parameter $\xi$ becomes sufficiently high as
\begin{equation}
\xi \equiv  L_{\rm ill} /n_{\rm e} r^2 > 10^3 .
\label{eq:xi_1}
\end{equation}
Here,
$L_{\rm ill} \simeq 1 \times 10^{36}$~erg~s$^{-1}$ is the illuminating
luminosity above $E_{\rm K}=27$~keV, and $r$ is 
the distance from the illuminating source.
Although this $r$ should be generally larger than $R_{\rm NS}$,
the hard X-ray source, i.e., the hot inflowing corona,
is considered to surround the NS rather spherically,
and subtend a significant solid angle as seen from each location of the atmosphere.
Therefore, $r \sim R_{\rm NS}$ is thought to be a safe approximation. Then, equation (\ref{eq:ne}) finally reduces to
\begin{equation}
n_{\rm e} = 48 n_{\rm x} + n_{\rm H}< 1 \times 10^{21}~{\rm cm}^{-3}~.
\label{eq:17}
\end{equation}
In figure~\ref{fig:conditions}, the region allowed by this constraint is indicated by a blue shaded rectangle.
We confirm that the red line does intersect this region,
where the Cd ions have $10^7$ to $10^8$ times the solar abundance.

\subsection{Visibility of the \texttt{redge} photons}

Supposing that the \texttt{redge} photons are emitted 
at a certain height above the NS surface, 
the atmosphere above the emission zone needs to be sufficiently transparent. 
This can be stated in terms of two optical depths. 
One is that the Thomson optical depth (i.e., the photon-scattering by free electrons) above the emission region must be less than 1 
in order for the \texttt{redge} photons to reach us. 
The other is that the emitted photons, which mostly have energies slightly above $E_{\rm K}$, should not be photoelectrically re-absorbed by ions of the same element that lie in higher atmosphere and remain in Hydrogen-like or lower ionization states. 

\subsubsection{Thomson optical depth}
\label{subsubsec:Thomson_optical_depth}

Using the Thomson cross section  $\sigma_{\rm T} = 6.65 \times 10^{-25}$~cm$^{2}$
and the atmospheric scale height $H$,
the first of the above two requirements
can be expressed as
\begin{equation}
\sigma_{\rm T} \; n_{\rm e} \; H <1  ~~\rightarrow ~~
n_{\rm e} H < 1.5 \times 10^{24} ~ {\rm cm^{-2}}~.
\label{eq:tau_Thomson_1}
\end{equation}
Using equation~(\ref{eq:ne}) and equation~(\ref{eq:scaleheight}),
this condition can be rewritten as
\begin{equation}
(48 n_{\rm x}+n_{\rm H}) (n_{\rm x}+n_{\rm H})/ (A_{\rm x} n_{\rm x}+n_{\rm H})
            < 2 \times 10^{22} ~{\rm cm^{-3}}~.
\label{eq:19}
\end{equation}
%
In figure~\ref{fig:conditions},
the region that satisfies this condition is 
bounded by a magenta rectangle-like shape, which is approximated as
$n_{\rm H} =2 \times 10^{22} ~{\rm cm^{-3}}$
at $n_{x} = 0$ and $n_{\rm x} =5 \times 10^{22} ~{\rm cm^{-3}}$.
It fully contains that for the Cd/In ionization.

\subsubsection{Self-absorption}
\label{subsubsec:Self_absorption}

Even when equation (\ref{eq:17}) is satisfied, 
a certain fraction of the Cd or In ions would remain in a H-like state 
instead of complete ionization, 
and their remaining K-shell electrons can absorb the recombination photons. 
To prevent this self-absorption, 
$n_{\rm x}$ needs to satisfy the condition as 
\begin{equation}
\tau_{\rm pe} = n_{\rm x} \sigma_{\rm pe} H < 1, \label{eq:tau_pe}
\end{equation}
where $\sigma_{\rm pe} = 3 \times 10^{-21}~{\rm cm^2}$ 
is the photoelectric absorption coefficient for $Z = 48$ \citep{Noguch_1987}. 
Using equation~(\ref{eq:scaleheight}), 
equation~(\ref{eq:tau_pe}) can be rewritten as 
\begin{equation}
\frac{n_{\rm x} (n_{\rm x} + n_{\rm H})}{A_{\rm x} n_{\rm x} + n_{\rm H}} < 4 \times 10^{18} ~{\rm cm^{-3}} ~. 
\label{eq:21 condition self abs}
\end{equation}
In figure~\ref{fig:conditions}, this condition is expressed 
by a black curve and grey shades. Thus, we confirm that 
the overlapping range between the red line and the blue rectangle
in figure ~\ref{fig:conditions}, as identified in 
section~\ref{subsubsec:vol_emission_measure},
survives the two optically-thin requirements.

\subsubsection{Additional requirements}
\label{subsubsec:dditional}

In addition to the above two constraints, 
we may need to consider two more requirements.
One is that the ionizing hard X-rays must reach the reprocessing zone
in the NS atmosphere without being significantly attenuated. This is automatically fulfilled
when the constraints considered in section~\ref{subsubsec:Thomson_optical_depth}
and section~\ref{subsubsec:Self_absorption} are both satisfied.

The other is that the Comptonizing corona,
which possibly intervenes between the NS and us,
should have a low Thomson optical depth,
so that the \texttt{redge} photons can freely get through it.
As discussed in section 4.2 of Paper I, the coronal optical depth in the present observation was relatively low, $\tau \sim 1.0$,
as specified by the $\nu F \nu$ spectrum which is nearly flat in the 3--20 keV range (Figure 1 of Paper I). Therefore, this condition is satisfied. An interesting case is another \textit{Suzaku} data set
of Aql X-1 acquired in 2011. On this occasion, the continuum was considerably harder, indicating $\tau > 3 $, and the 30 keV feature was not visible (Paper I).

\medskip
\section{Comparison with Nucleosynthesis calculations for X-ray bursts}
\label{sec:xrbcalc}

In section~\ref{section:constraints}, 
we found that the 30 keV feature can be interpreted
as the \texttt{redge} quasi-continuum from Cd/In isotopes,
provided that $n_{\rm x}$ and $n_{\rm H}$ lie on the intersection 
between the red line and the blue region in figure~\ref{fig:conditions}. The most stringent requirement set by this intersection is that the $n_{\rm x}/n_{\rm H}$ ratio should be high enough, as
\begin{eqnarray}
    \frac{n_{\rm x}}{n_{\rm H}} \gtrsim 8\times10^{-4}~
    \Longleftrightarrow 
    X_{\rm x} \gtrsim 0.06\left(\frac{A_{x}}{100}\right)\left(\frac{X_{\rm H}}{0.73}\right). \label{eq:const1}
\end{eqnarray}
Here,  $X_{\rm x}$ is the mass fraction of the Cd/In isotopes, 
and $X_{\rm H}$ is that of hydrogen. Hereafter, we focus on this requirement.

In the present scenario, the Cd/In ions are thought to come 
mainly via decays of parent nuclei 
that should be synthesized in the  \textit{rp}-process 
during the phase of X-ray burst activity.
Therefore, when trying to identify the relevant isotopes, the \textit{real} elapsed time between the present observations (D2--D4) and the immediately preceding X-ray burst 
is particularly important.

According to the burst catalog MINBAR~\citep{galloway20}, 
the last X-ray burst from Aql X-1 before the present observations 
was detected by the RXTE/PCA
on 2007 September 22, which roughly corresponds to D1. 
Thus, we assume that the maximum elapsed time is about 16.2 days, 
which is the separation between D1 and D4. 
However, missing several X-ray bursts may have occurred in this period,
because none of the instruments (RXTE/PCA, BeppoSAX/WFC, and INTEGRAL/JEM-X) 
that comprise the MINBER catalog have genuine all-sky coverage. 
Considering these, we might think of the following two cases 
for the detection of the \texttt{redge} feature:\\
(i) There was no missing burst, and hence the elapsed time is longer than a few hours. In this case, the relevant isotopes must be
stable on a timescale of $\gtrsim 1$ day.\\
(ii) The present observations D2-D4 all took place each immediately after an X-ray burst, 
with an elapsed time of an hour or less. 
In this case, several short-lived nuclides also become candidates.

In this paper, we mainly consider the case (i), 
conservatively assuming that no burst occurred
during the \textit{Suzaku} observations.
This assumption is well justified in view of the $\alpha$ parameter, which represents the ratio between the time-averaged
persistent and burst energy fluxes. When the burst peak flux of an XRB source is identified with the Eddington flux, its burst recurrence time $\Delta t$ is expressed as
\begin{eqnarray}
    \Delta t \sim \left(\dot{M}/\dot{M}_{\rm Edd}\right)^{-1}\alpha\tau~,\label{eq:alpha}
\end{eqnarray}
where $\tau$ is the burst duration. Using $\alpha\approx40$ and  $\tau\approx15$-$40~{\rm s}$ which are typical of hydrogen-rich bursts (e.g., \cite{Heger2007}), 
and the Eddington-normalized accretion rate $\dot{M}/\dot{M}_{\rm Edd}\sim0.01$ observed by \citet{Sakurai12} from Aql X-1,
we obtain $\Delta t\sim60$-$160~{\rm ks}$. This is very close to the actual elapsed time from D1 to D4, 66.5 ks.

\begin{figure}[htb]
\begin{center}
\includegraphics[width=\linewidth]{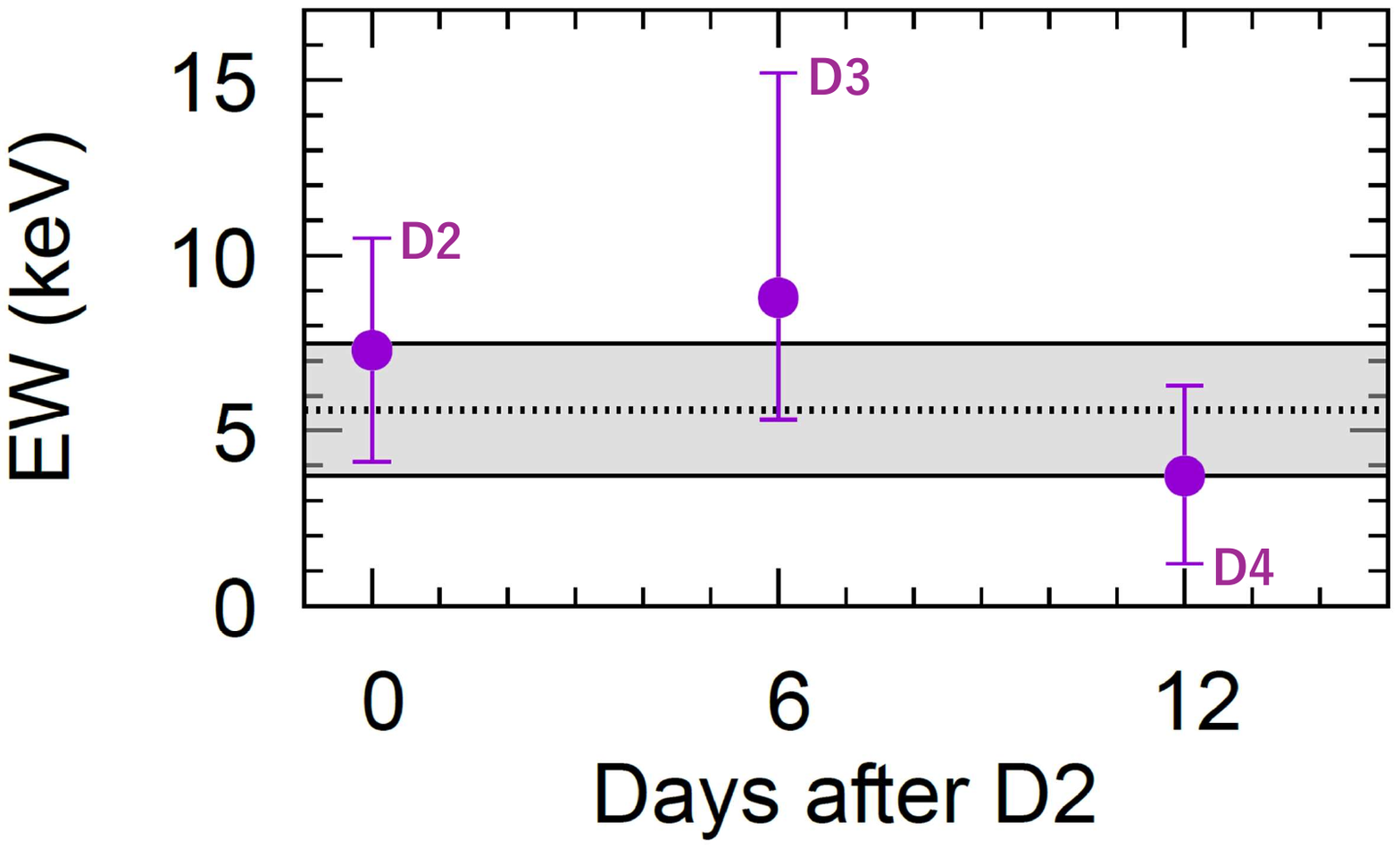}
\end{center}
\caption{The EWs of the 30-keV feature, 
 determined with the Gaussian fitting (Paper I) in D2, D3, and D4, 
 plotted as a function of time (days) after D2. 
 The dotted line and gray region, respectively, indicate the weighted mean 
 and the standard deviation, 
 as $ \overline{\rm EW}=5.6\pm1.9~{\rm keV}$.
}
\label{fig:possible decay}
\end{figure}

Then, to satisfy the condition of equation (\ref{eq:const1}), the synthesized Cd/In isotope must have a very long half-life time, 
or be a stable nuclide. 
This conjecture would also be supported by the evolution of EW, which is presented in figure~\ref{fig:possible decay}.
We can see that the 30 keV hump is consistent 
with being constant from D2 to D4. It is therefore suggested that the same Cd/In ions kept producing the 30 keV hump from D2 to D4.
If it were not a stable or long-lived nuclide, we would need to assume that three missing bursts occurred each just before the three observations, which seems unlikely.

  To examine whether Cd/In can contribute more than
  6\% of the total mass  to satisfy equation~(\ref{eq:const1}), 
  we performed a one-zone model calculation of X-ray bursts.
  in the same way as \citet{Dohi20},
  where most of reaction rates are taken 
  from the latest Jina Reaclib data~\citep{C10}
  \footnote{They are updated on 2021 June 24. 
  See \url{https://reaclib.jinaweb.org/library.php?action=viewsnapshots}}.   
  
  Our one-zone model has essentially three parameters;
  the ignition depth $y_{\rm ign}$ ($=P_{\rm ign}/g$ 
  where $P_{\rm ign}$ is the ignition pressure), 
  the initial mass fraction of hydrogen relative to helium, $X_{\rm H}/X_{\rm He}$, 
  and the initial CNO metallicity $Z_{\rm CNO}$. 
  We fix $X_{\rm H}/X_{\rm He}=2.9$ and the initial metal compositions 
  as 35\% of $^{14}{\rm O}$ and 65\% of $^{15}{\rm O}$~\citep{Hanawa83}, 
  which realizes the condition that SnSbTe-group nuclei are abundant~\citep{Koike04,2025PhRvC.112f5808Z}.

\begin{figure}
    \centering
    \includegraphics[width=\linewidth]{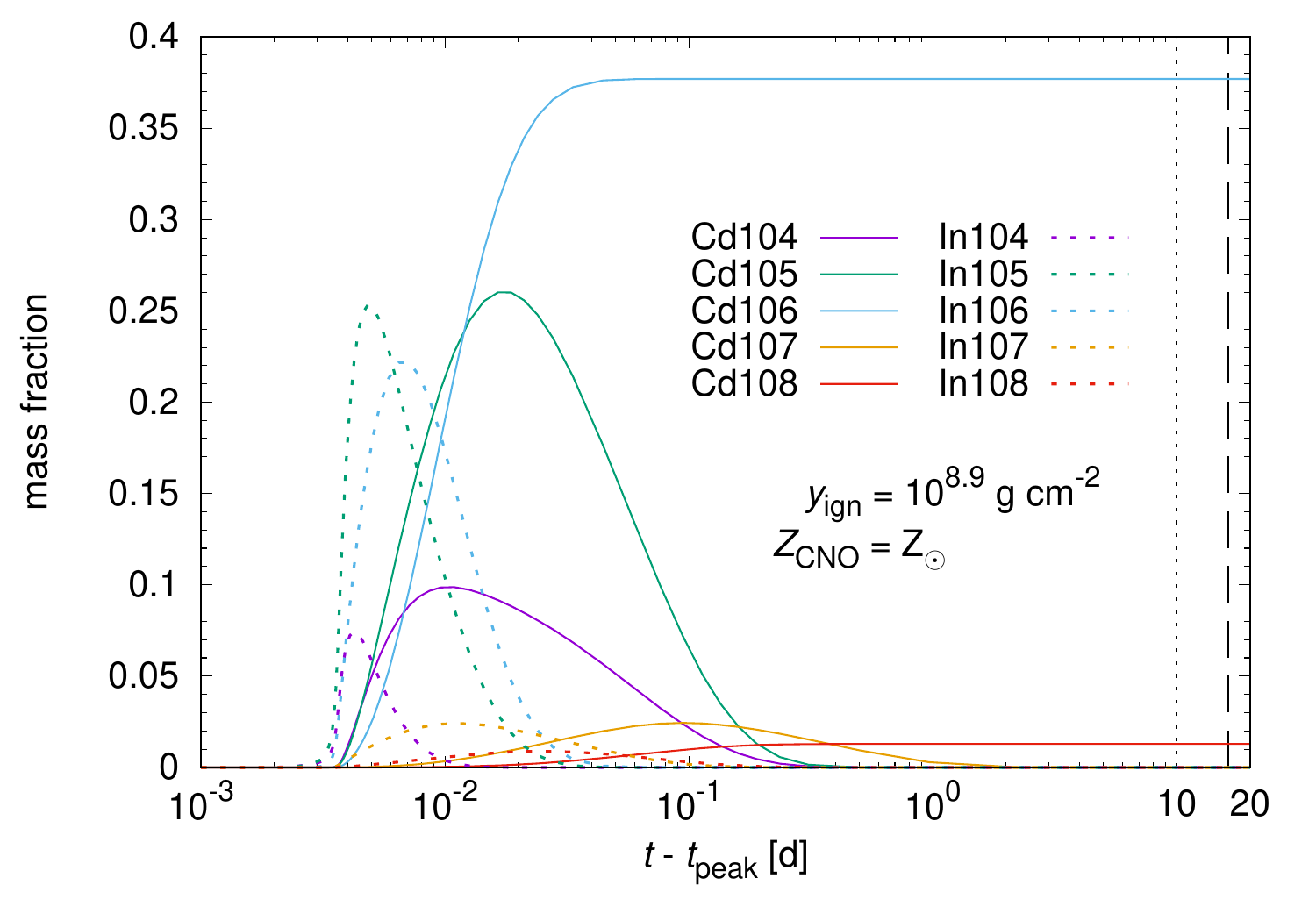}
    \caption{Evolution of mass fraction of several \textit{rp}-process nuclei. The abscissa is the elapsed time after $t_{\rm peak}$, which is the time when the specific luminosity becomes maximum. The dotted and dashed lines indicate the timings of D2 and D3, respectively. Other isotopes that are not shown are little produced ($<1\%$).}
    \label{fig:fmass}
\end{figure}

Figure~\ref{fig:fmass} presents the evolution of several isotopes of Cd and In. Here, we chose the initial CNO abundance to be solar, and $y_{\rm ign}=10^{8.9}~{\rm g~cm^{-2}}$. The latter yielded the highest Cd/In abundances among the parameter regions we examined. As we see, $^{106}{\rm Cd}$ and $^{108}{\rm Cd}$ can persist for  $\gtrsim 10$ days since they are stable nuclei. In particular, $^{106}{\rm Cd}$ becomes abundant in this case with the mass fraction of 38\% (see also similar result reported by \cite{Koike04}). 
This is because $^{106}{\rm Cd}$ is synthesized from $\beta^+$-decay chains of proton-rich $A=106$ isobars 
(such as ${}^{106}{\rm Te}$ and ${}^{106}{\rm Sb}$) that are involved in the abundant SnSbTe-cycle nuclide~\citep{Schatz01}.

\begin{figure}
    \centering
    \includegraphics[width=\linewidth]{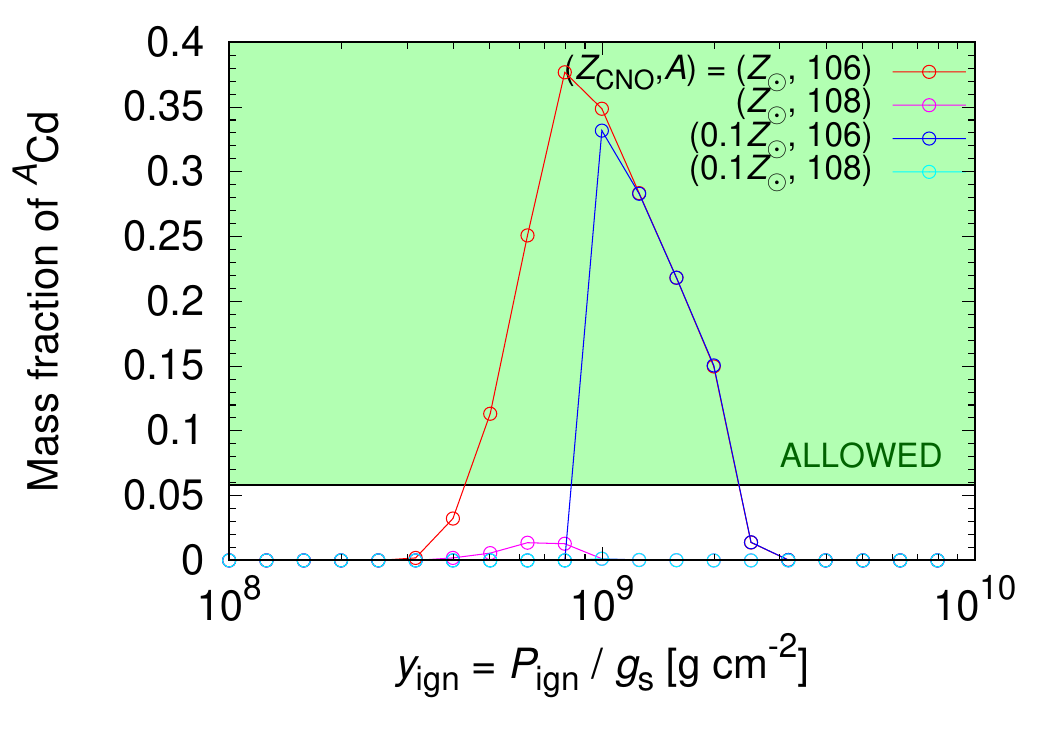}
    \caption{The calculated mass fractions of ${}^{106}{\rm Cd}$ (red/blue) and ${}^{108}{\rm Cd}$ (magenta/cyan), 
    10 days after a thermonuclear burst, shown against the assumed ignition depth. The assumed CNO abundance is 1 Solar ($Z_{\rm CNO}=Z_\odot$), or 0.1 Solar ($Z_{\rm CNO}=0.1 Z_\odot$). The region satisfying equation (\ref{eq:const1}) is shown in green.
    }
    \label{fig:cd}
\end{figure}

The final mass fractions of 
${}^{106}{\rm Cd}$ and ${}^{108}{\rm Cd}$, at 10 days after the ignition\footnote{These mass fractions have already converged at $t\sim 0.1~{\rm day}$}., 
are shown in figure~\ref{fig:cd} as a function of $y_{\rm ign}$.
In a wide parameter region over $\log_{10} (y_{\rm ign}/[{\rm g~cm^{-2}}])$=8.7--9.3, we can see that ${}^{106}{\rm Cd}$ is produced plenty enough to account for the observational constraint
expressed by equation~(\ref{eq:const1}). 
On the other hand, ${}^{108}{\rm Cd}$ is synthesized at most about 2\%, 
and cannot be a candidate for the observed hump feature. 

How about other candidates for the observed hump feature? 
Figure~\ref{fig:fmass} also shows the time evolution of other Cd and In isotopes, 
which might produce the \texttt{redge} emission if the scenario (ii) holds. 
For $t\gtrsim0.2~{\rm d}$ after the X-ray burst, however, they finally decay to stable nuclei, such as Ag and Pd, because of their short half-lives. We also examined the production of more neutron-rich Cd or In isotopes with $A\ge109$, because the \textit{rp}-process could technically make their parent nuclei heavier 
than the SnSbTe-group nuclei up to $^{126}{\rm Xe}$~\citep{Koike04}. 
However, we found that none of them are produced more than $\sim0.1$\% at least under the parameter ranges we considered. Thus, we conclude that ${}^{106}{\rm Cd}$ provides, in the \textit{rp}-process scenario, the most likely origin of the 30 keV feature observed from Aql X-1.

\section{How to lift up the Cd ion?}
\label{sec:further_issues}

In sections \ref{section:constraints} and \ref{sec:xrbcalc}, 
we presented a simple theoretical calculation 
which supports the observational inference 
on the richness of Cd or In isotopes 
due to the strong \textit{rp}-process in the atmosphere of Aql X-1. 
Nevertheless, we are still left with an important issue 
that cannot be answered within the one-zone model framework. 
Namely, how the \textit{rp}-process ashes can be lifted up to the atmosphere. 
This problem is examined below.

Although we have successfully identified the physical conditions
of the emission region (figure~\ref{fig:conditions}), the actual height above the NS surface remains unspecified, because the atmospheric density at the NS surface remains unknown. In any case, it would be at considerable heights above the NS surface.
If so, the implicit assumption that we have adopted so far, that the heavy-element ions have the same height distribution as protons and electrons, would not necessarily be justified. In reality, heavier ions must have much lower scale heights than less massive particles, assuming a local thermal equilibrium among different particle species. We therefore need to consider mechanisms that counteract the gravitational pull of heavy elements, allowing them to rise to higher zones of the atmosphere.

At the onset of an X-ray burst, the thermonuclear flash triggers the development of a convective region. In fact, \cite{2024ApJ...975..250G} showed that, in the hydrogen-rich accretion case, the convection zone encompasses the burst ignition layer within $\sim0.1~{\rm s}$ after the onset of a thermonuclear helium flash. Since heavy \textit{rp}-process elements are the most abundant around such a \textit{deep} ignition zone, they are temporarily lifted up to the atmosphere regions. However, this convection triggered by the helium flush ends in a few seconds~\citep{2004ApJS..151...75W}. To realize our present scenario for heavy ions being kept in the atmosphere, it is necessary to invoke an additional mechanism not related to the thermonuclear burning, which is discussed below.

The most likely mechanism for the heavy-ion lift-up is 
strong atmospheric convection driven by turbulent mixing of matter, accompanied by several hydrodynamical instabilities~
(see \cite{1988A&A...198..163F,1993ApJ...419..768F} for details). Generally, the atmosphere of accreting NSs differentially rotates
in both radial and latitudinal directions,
due to accretion-induced angular momentum transport~
\citep{1999AstL...25..269I}. 
Then, unless the accretion rate is too low, such a differential rotation can induce the toroidal magnetic field up to $\sim10^{10}~{\rm G}$. 
Depending on the accretion rate and neutron-star spin, the toroidal field could become subject to a non-axisymmetric ($m=1$ mode) Taylor instability, 
which generates a radial (and poloidal) magnetic field 
(e.g., \cite{2007A&A...474..145Z,2019ApJ...881...66G} 
for non-degenerate stars).

This instability, called Taylor-Spruit instability \citep{2002A&A...381..923S}, is triggered by the turbulent mixing at high accretion rates, because the generated radial magnetic field, 
which is usually frozen into the plasma, 
can provide magnetic buoyancy,
and hence spread matter over the atmosphere. Since the magnetic vs. gravitational force ratio working on Cd ions is only by a factor $\sim 2$ lower than that for protons, not only light elements such as hydrogen and helium, but also heavy elements (Cd/In) would be lifted in the atmosphere high enough to be observed as the recombination continuum source. According to \cite{pb07}, this magnetic turbulent mixing works throughout the atmosphere unless the accretion rate is too low.

Along with the above consideration, let us more quantitatively examine the scenario of lifting the Cd ions, focusing on two aspects.
One is whether the accretion rate, $\dot{M}$, exceeds the critical value that enables the Taylor-Spruit instability.
By slightly modifying equation (59) of \cite{pb07},
this condition is given as
\begin{eqnarray}
    \frac{\dot{M}}{\dot{M}_{\rm edd}}&\gtrsim&5.5\times10^{-3}\alpha_{\rm TS}^{-3}\rho_5T_8\left(\frac{1.33}{\mu}\right)^{2}\left(\frac{\Delta \ln \mu}{0.48}\right)
 \nonumber \\
    &&\left(\frac{\Omega}{550~{\rm Hz}}\right)^3\left(\frac{R_{\rm NS}}{12~{\rm km}}\right)^{-2}\left(\frac{g_{\rm eff,14}}{0.55\times1.6}\right)^3~. \label{eq:mdotc}.
\end{eqnarray}
Here, $\dot{M}_{\rm edd}=1.3\times10^{18}~{\rm g~s^{-1}}$ 
is the Eddington accretion rate 
(for $M_{\rm NS}=1.4~M_\odot, R_{\rm NS}=12~{\rm km}$ and $X=0.73$);
$\alpha_{\rm TS}$ is an uncertain factor of order unity on 
the mixing diffusivity in the Taylor-Spruit instability;
$\mu\approx1.33$ is the mean molecule weight assuming pure He; $\rho_5$ is the mass density in units of $10^{5}~{\rm g~cm^{-3}}$,
which roughly corresponds to the minimum density of the convective regions in the early phase of X-ray bursts \citep{2004ApJS..151...75W};
$T_8$ is the temperature in units of $10^8~{\rm K}$,
which is reasonable (or possibly slightly lower) 
for the hard state in  D2--D4 according to the inferred 
internal temperature of Aql X-1~(e.g., \cite{2021PhRvD.103f3009L});
$\Omega\approx550~{\rm Hz}$ is the rotational frequency of Aql X-1~
\citep{2008ApJ...674L..41C};
$\Delta \ln \mu\approx0.48$ is the logarithmic change in 
the mean molecular weight at the He/Cd boundary; $g_{{\rm eff},14}$ is the effective surface gravity 
in units of $10^{14}~{\rm cm~s^{-2}}$, 
including the electric-field force by $^{106}{\rm Cd}$ ions
\footnote{Assuming the presence of a single ionized 
heavy element species for simplicity, the effective gravity force is locally described as
\begin{eqnarray}
    F_{\rm eff} = A m_u g_{\rm eff}>0~~\rm{with}~~g_{\rm eff}\equiv g(1-Z/A)~,
\end{eqnarray}
where $m_u$ is the atomic mass constant. 
This implies that the radial electric fields owing to heavy ions cannot contribute to lifting them up, although they reduce the strength of gravity, making it easier to cause turbulent mixing.}. 
Meanwhile, \cite{Sakurai12} reported $\dot{M}/\dot{M}_{\rm edd}\approx 0.08$ and 0.01 in D1 and D3, respectively, latter of which is similar to those in D2 and D4 (see their Figure 2). These imply that the turbulent mixing is indeed possible.

The other issue is whether the \textit{rp}-process products, synthesized in the X-ray burst that occurred at about D1,
can really stay plenty enough for $\gtrsim10~{\rm days}$ in the convective atmosphere,
when considering the dilution effect 
by the continuous in-flow of fresh accreting matter. As already mentioned, the convective matter mixing can reach the ignition depth $y_{\rm ign}\sim10^{9}~{\rm cm~s^{-2}}$, where the Cd/In isotope is the most abundant as in figure \ref{fig:fmass}. 
Then, the total amount of the convective matter, 
consisting mostly of hydrogen and helium (98\% of total) just after the end of the X-ray burst (when we set $t=0$), is estimated under the one-zone approximation as
\begin{eqnarray}
    \Delta M_0\approx 4\pi R_{\rm NS}^2y_{\rm ign} = 2\times10^{23}~{\rm g}~. ~\label{eq:m0}
\end{eqnarray}
 On the other hand, the Cd/In masses in the convective layer at $t=0$ can be estimated as
\begin{eqnarray}
    \Delta M_{x,0}\simeq X_x(0)E_{\rm tot}/Q_{\rm nuc}\sim 10^{20}~{\rm g}
\end{eqnarray}
where $Q_{\rm nuc}\simeq1.6+4X_{\rm H}$ MeV nucleon$^{-1}$ is the nuclear energy release per nucleon during the X-ray burst (e.g., see \cite{2017PASA...34...19G}), $E_{\rm tot}\sim10^{40}~{\rm erg}$ is local total burst energy
\footnote{We note that one-zone models that meet the condition of Eq. (\ref{eq:const1}) show $E_{\rm tot}=$(1--4)$\times10^{40}~{\rm erg}$.}, and $X_x(t)$ is the Cd/In mass fraction in a time $t$ after the end of the X-ray burst (hence $X_x$ defined in Section \ref{sec:xrbcalc} becomes $X_x(0)$ in this notation). 

The Cd/In masses in the convective layer, $\Delta M_x(t)$, is given as
\begin{align}
    \Delta M_x(t) = 4\pi R_{\rm NS}^2 y_{\rm mix}(t) X_x(t)~. \label{eq:dmx}
\end{align}
Here, $\Delta M_x(0)=\Delta M_{x,0}=\Delta M_0X_x(0)$ and $y_{\rm mix}(t)$ indicates the column depth at the bottom of the convection layer. Starting from the ignition shell condition as
$y_{\rm mix}(0)=y_{\rm ign}$, the evolution of $y_{\rm mix}$ can be phenomenologically given as~(see Section 5 in \cite{pb07})
\begin{eqnarray}
    y_{\rm mix}(t) = y_{\rm ign}\left(1+\frac{\dot{M}t}{\Delta M_0}\right)^\alpha~, \label{eq:ymix}
\end{eqnarray}
where $\alpha$ denotes the characteristic index for the growth of the convective layer: 
$\alpha=0$ corresponds to no evolution of convective regions (i.e, $\Delta M_0$ is conserved) 
while $\alpha=1$ to its linear growth according to accretion. The numerical analysis of \cite{pb07} showed $\alpha\approx0.73$ for sufficient time, implying that the convective layer slowly develops compared to the accretion timescale. Then, from Eqs. (\ref{eq:m0}), (\ref{eq:dmx}) and (\ref{eq:ymix}), $X_x(t)$ is expressed as
\begin{eqnarray}
    X_x(t) = \frac{\Delta M_x(t)}{\Delta M_0}\left(1+\frac{\dot{M}t}{\Delta M_0}\right)^{-\alpha} \label{eq:dxx}
\end{eqnarray}

Assuming that the convective matter is homogeneously distributed and Cd/In ions in that matter could bury into only deeper radiative atmosphere regions 
due to the component weight\footnote{Strictly speaking, if it is not the case, 
the characteristic dilution timescale could become longer for the small $\alpha$. 
However, this does not affect the comparison of timescales in Eq. (\ref{eq:ts}).}, the convective mass loss of Cd/In can be written as
\begin{eqnarray}
    \frac{d\Delta M_x(t)}{dt} = -\dot{M}X_x(t) = -\frac{\dot{M}\Delta M_x(t)}{\Delta M_0}\left(1+\frac{\dot{M}t}{\Delta M_0}\right)^{-\alpha}~,
    \label{eq:diff_mx}
\end{eqnarray}
where the second equality follows from Eq. (\ref{eq:dxx}). Since no Cd/In isotopes are synthesized in scenario (i) after the end of the X-ray burst, the Cd/In masses in convective layers decrease monotonically with time. 

By solving Eq. (\ref{eq:diff_mx}), we obtain
\begin{widetext}
\begin{eqnarray}
    \frac{\Delta M_x(t)}{\Delta M_{x,0}} = \frac{y_{\rm mix}(t)X_x(t)}{y_{\rm ign} X_{x}(0)} = 
    \begin{cases}
       \exp\left(-\frac{\dot{M}t}{\Delta M_0}\right) & \alpha=0  \\
        \exp\left[-\frac{1}{1-\alpha}\left\{\left(1+\frac{\dot{M}t}{\Delta M_0}\right)^{1-\alpha}-1\right\}\right] & 0<\alpha<1 \\
        \left(1+\frac{\dot{M}t}{\Delta M_0}\right)^{-1} & \alpha=1  ~.
        \end{cases}
        \label{eq:sol}
\end{eqnarray}
\end{widetext}
This solution has the following two effects on the mass loss of Cd/In ions. One is burial of the Cd/In ions from the convective zone into the deeper radiative zone. The other is a deepening of the convective layers through accretion, diluting Cd/In convective matter. 
These are explicitly represented by
$\alpha=0$ and 1 in Eq. (\ref{eq:sol}), respectively, 
which show exponential and power-law decays.  

An important point is that both cases result in 
the same characteristic decay timescale given as
\begin{eqnarray}
    \tau\equiv {\Delta}M_0/\dot{M}\sim 10^7~{\rm s} >  10~{\rm day}~, \label{eq:ts}
\end{eqnarray}
where $\dot{M}\sim0.01\dot{M}_{\rm Edd}=1.3\times10^{16}~{\rm g~s^{-1}}$ 
is the accretion rate corresponding to D2-D4. 
Therefore, the Cd/In can remain in the convective atmosphere for more than 10 days, regardless of $\alpha$ values.
From the conditions of turbulent mixing in Eqs. (\ref{eq:mdotc}) and (\ref{eq:ts}), 
we conclude that our scenario is feasible.

It should be mentioned that the above discussion still depends on the quantitative properties of the turbulent mixing in accreting neutron stars. For example, the operation of the Taylor-Spruit instability, 
characterized with $\alpha_{\rm TS}$, 
would be sensitive to the toroidal magnetic-field configurations and fluid microphysics \citep{2024ApJ...974..290S}. Also, how the convective regions spread and how the minimum density therein is determined should depend strongly on the actual performance of X-ray bursts. 
To reduce these uncertainties, it may be worth performing a global X-ray burst simulation, including the lifting-up mechanism, and the multi-dimensional effects such as flame propagation and fast rotation~(\cite{2025arXiv250513879J} and reference therein). However, further attempts on these issues are clearly beyond the scope of the present paper.



\section{Conclusion}
\label{section:conc}
In Paper I, we described an enigmatic spectral hump of Aquila X-1,
seen by \textit{Suzaku} at $\sim 30$ keV of its hard-state spectra
during its 2007 outburst.
The hump was tentatively identified with a \textit{redge} 
quasi-continuum from highly ionized Cd or In isotopes
that were produced in X-ray bursts via the \textit{rp}-process.
In the present paper, we examined the feasibility of this scenario
from various aspects, and obtained the following results.
\begin{enumerate}
\item 
The \textit{redge} interpretation was reinforced, 
by examining the spectrum for the cascade K$\alpha$ line
which should accompany the recombination-edge feature.
\item
We quantified the physical conditions of the neutron-star atmosphere
that enable the \textit{Suzaku} detection of the \textit{redge} feature.
The derived major constraint is that Cd/In in the relevant region
must have 7--8 orders of magnitude higher abundance than the Solar value.
\item 
Inspection of an X-ray burst catalog indicated 
that the present \textit{Suzaku} observations are likely 
to have occurred about 10 days after the last preceding X-ray burst.
\item 
Through a one-zone model calculation which can produce SnSbTe-group nuclei due to the strong \textit{rp}-process, we identified $^{106}$Cd as the most promising candidate isotope,
because it is a stable nuclide, and produced plenty enough to apparently satisfy the able requirement. The present results thus potentially provide direct observational evidence for the \textit{rp}-process operation in X-ray bursts.
%
\item There still remain two issues to be solved.
One is how to lift up the heavy ions up to certain heights in the atmosphere, and the other is how to prevent them from being diluted by the fresh accreting materials. As a solution to these issues,
we considered turbulent matter mixing in surface regions of accreting neutron stars. We found this scenario possible from minimum astrophysical aspects, although quantitative uncertainties relevant to the dynamics of the lifting-up mechanism are large.
\end{enumerate}

\begin{acknowledgments}
    We are grateful to N. Nishimura for providing the impetus for this collaboration. This work is supported by JSPS KAKENHI (25K17403).
\end{acknowledgments}

\bibliography{ref}{}
\bibliographystyle{aasjournal}

\end{document}